# Flexible heat-flow sensing sheets based on the longitudinal spin Seebeck effect using one-dimensional spin-current conducting films

Akihiro Kirihara[1,2*], Koichi Kondo[3], Masahiko Ishida[1,2], Kazuki Ihara[1,2], Yuma Iwasaki[1], Hiroko Someya[1,2], Asuka Matsuba[1], Ken-ichi Uchida[4,5], Eiji Saitoh[2,4,6,7], Naoharu Yamamoto[3], Shigeru Kohmoto[1] and Tomoo Murakami[1]

[1]*Smart Energy Research Laboratories, NEC Corporation, Tsukuba, 305-8501, Japan*

[2]*Spin Quantum Rectification Project, ERATO, Japan Science and Technology Agency, Sendai, 980-8577, Japan*

[3]*NEC TOKIN Corporation, Sendai, 982-8510, Japan*

[4]*Institute for Materials Research, Tohoku University, Sendai, 980-8577, Japan*

[5]*PRESTO, Japan Science and Technology Agency, Saitama, 332-0012, Japan*

[6]*WPI, Advanced Institute for Materials Research, Tohoku University, Sendai, 980-8577, Japan*

[7]*Advanced Science Research Center, Japan Atomic Energy Agency, Tokai, 319-1195, Japan*

* e-mail: a-kirihara@cw.jp.nec.com

**Heat-flow sensing is expected to be an important technological component of smart thermal management in the future. Conventionally, the thermoelectric (TE) conversion technique, which is based on the Seebeck effect, has been used to measure a heat flow by converting the flow into electric voltage. However, for ubiquitous heat-flow visualization, thin and flexible sensors with extremely low thermal resistance are highly desired. Recently, another type of TE effect, the longitudinal spin Seebeck effect (LSSE), has aroused great interest because the LSSE potentially offers favourable features for TE applications such as simple thin-film device structures. Here we demonstrate an LSSE-based flexible TE sheet that is especially suitable for a heat-flow sensing application. This TE sheet contained a $Ni_{0.2}Zn_{0.3}Fe_{2.5}O_4$ film which was formed on a flexible plastic sheet using a spray-coating method known as "ferrite plating". The experimental results suggest that the ferrite-plated film, which has a columnar crystal structure aligned perpendicular to the film plane, functions as a unique one-dimensional spin-current conductor suitable for bendable LSSE-based sensors. This newly developed thin TE sheet may be attached to differently shaped heat sources without obstructing an innate heat flux, paving the way to versatile heat-flow measurements and management.**

As efficient energy utilization is becoming a crucially important issue for sustainable future, a heat management technique to optimally control the flow of omnipresent thermal energy is currently of great interest. To realize smart thermal management with real-time controllability, there has been a growing demand for visualizing the flow of heat in various places such as industrial facilities and large-scale data centres. The thermoelectric (TE) conversion technique [1-3], which directly converts a thermal gradient into an electric current, is one of the most powerful methods utilized to sense a heat flow as a voltage signal. In fact, heat-flow sensors based on the Seebeck effect [4], which have thermopile structures consisting of π-structured thermocouples, are commercially available and used for various purposes such as the evaluation of materials.

To further extend heat-flow-sensing capabilities to other widespread applications, however, such conventional devices face certain challenges. First, because Seebeck-based TE devices exhibit a relatively high heat resistance, the introduction of these devices into a heat-flow environment inevitably obstructs the heat flux and alters the distribution of the heat flow. Therefore, it is difficult to correctly evaluate the innate heat flux which we actually want to determine. Second, most of the commercially available heat-flow sensors are rigid and not easily applied to curved or uneven surfaces, making it difficult to monitor the heat flux around irregularly shaped heat sources. Because conventional TE devices, in which thermocouples are connected electrically in series, are intrinsically vulnerable to bending stresses, materials and structures for flexible TE devices have been extensively studied [5-7].

For such sensing applications, an emerging research field, spin caloritronics [8,9], will provide new device-design opportunities. For example, TE devices based on the anomalous Nernst effect (ANE), which exhibit transverse TE voltage in ferromagnetic metals (FM), can be suitably utilized for sensing purposes [9-13]. In this work, we present another promising approach to realizing flexible heat flow sensors using the longitudinal spin Seebeck effect (LSSE) [14-18]. First reported in 2010, the LSSE offers an unconventional method to design TE devices by making use of a physical quantity called a spin current. The LSSE devices, typically composed of a ferromagnetic insulator (FI) and a normal metallic film (NM), have gained attention because of the simple device structure and novel scaling capability, leading to novel TE devices [16]. The LSSE also has potential to realize practical sensing applications. It is recently reported that $(FI/NM)_n$ multilayer structure unexpectedly exhibit significantly enhanced LSSE signal [19], which may lead to high-sensitive heat-flow sensors. Furthermore, combination of the LSSE and ANE is also a quite hopeful approach. In hybrid TE devices consisting of FI and FM layers, both the LSSE and ANE can constructively contribute to the output voltage, leading to largely enhanced TE signals [20,21]. To pave the way for practical sensing applications using the LSSE, here we have demonstrated LSSE-based heat-flow sensing sheets.

The concept of the LSSE-based flexible TE sheet is schematically depicted in Fig. 1. The sheet consists of a magnetic (ferro- or ferrimagnetic) film with in-plane magnetization **M** and a metallic film formed on a flexible substrate. When a heat flux $q$ flows through the TE sheet perpendicularly to the film plane, a spin current density $\mathbf{j}_s$ is induced by $q$ via the LSSE. The value of $\mathbf{j}_s$ is proportional to $q$ ($|\mathbf{j}_s| \propto q$). Then, $\mathbf{j}_s$ is converted into an electric field in the transverse direction via the inverse spin Hall effect (ISHE) [22,23]:

$$\mathbf{E}_{\text{ISHE}} = (\theta_{SH}\,\rho)\,\mathbf{j}_s \times \frac{\mathbf{M}}{|\mathbf{M}|} \quad (1).$$

In the above equation, $\theta_{SH}$ and $\rho$ represent the spin-Hall angle and resistivity of the metallic film, respectively. Therefore, the voltage signal $V$ between the two ends of the TE sheet can be employed to evaluate the heat flux $q$ penetrating through the sheet, because $V$ is proportional to $q$ ($V = E_{\text{ISHE}}\, l \propto q$). Here, it should also be emphasized that a longer sheet length $l$ straightforwardly leads to larger output voltage $V$. This scaling law is in stark contrast to that of conventional TE devices, in which TE voltage scales with the number of thermocouples connected within the devices. These features enable us to design simple bilayer-structured devices suitable for heat-flow sensors.

But there is a problem when we use the above setup for broad heat-flow sensing purposes. When $q$ flows obliquely to the TE sheet, the in-plain component of $q$ gives rise to another TE effect called the transverse spin Seebeck effect (TSSE) [24-28], which can also contribute to the output voltage. Since the mixed output signals from the LSSE and TSSE cannot be distinguished from each other, the TSSE becomes an encumbrance to the correct evaluation of $q$ penetrating the TE sheet in this case. To exclude the TSSE contribution, here we used unique one-dimensional (1D) spin-current conductor, which enables us to detect only the LSSE contribution and to correctly evaluate $q$ flowing across the TE sheet.

**Results**

**Fabrication of the LSSE-based TE sheet**   To demonstrate such an LSSE-based flexible TE sheet, we used a spray-coating technique known as "ferrite plating" to grow ferrimagnetic films. Ferrites refer to oxide ceramics containing iron, which typically exhibit ferromagnetic properties and have been successfully used as magnetic materials for LSSE devices [29,30]. However, conventional ferrite-film preparation techniques, such as liquid phase epitaxy and pulsed-laser deposition, require a high temperature process (400 – 800 °C) for crystallizing the ferrites, hindering the formation of films on soft-surfaced materials, such as plastics. By contrast, ferrite plating is based on a chemical reaction process in which the Fe ion is oxidized ($Fe^{2+} \rightarrow Fe^{3+}$); therefore, no high-temperature processes, such as annealing, are required [31,32]. This feature enables us to coat ferrite films on a variety of substrates, including plastic films.

In this work, we prepared a ferrite $Ni_{0.2}Zn_{0.3}Fe_{2.5}O_4$ film using this method. As schematically illustrated in Fig. 2(a), we grew the film by simultaneously spraying an aqueous reaction solution ($FeCl_2+NiCl_2+ZnCl_2$) and an oxidizer ($NaNO_2+CH_3COONH_4$) onto a substrate. In this process, the oxidizer oxidizes the chlorides in the reaction solution, forming a $Ni_{0.2}Zn_{0.3}Fe_{2.5}O_4$ film on the substrate. All the processes were performed below 100 °C.

A noticeable feature of the ferrite film, grown via such a layer-by-layer chemical process, was its columnar-crystal grain structure. Figure 2(b) depicts the cross-sectional scanning electron microscope (SEM) image of a $Ni_{0.2}Zn_{0.3}Fe_{2.5}O_4$ film that was grown on a $SiO_2$/Si substrate for the purpose of the SEM observation. The diameter of the columnar grain was typically approximately 100 nm. We also verified via transmission electron microscopy and electron diffraction measurements that the crystal orientation of the $Ni_{0.2}Zn_{0.3}Fe_{2.5}O_4$ was coherently aligned within a single columnar grain. Such a columnar structure can function as a 1D spin-current conductor favorable for LSSE-based (and TSSE-free) heat-flow sensors because of the following two reasons. First, in the LSSE configuration shown in Fig. 1, a magnon spin current is driven along the columnar grain and is thus less subject to grain scattering, effectively leading to the LSSE signal. Second, since the columnar-grain boundaries impede the transverse propagation of both magnons and phonons, in-plane components of a heat flow cannot effectively produce the TSSE in the light of previous studies (e.g., see ref. [33,34]). Thus we can exclude the possible TSSE contribution, enabling us to correctly measure a heat flow penetrating the TE sheet via the LSSE.

Using the ferrite plating technique, we successfully fabricated a flexible TE sheet based on the LSSE. Fig. 2(c) represents a photograph of the prepared TE sheet. First, a 500-nm-thick $Ni_{0.2}Zn_{0.3}Fe_{2.5}O_4$ film was grown on a 25-μm-thick polyimide substrate. Then, a Pt film with a thickness of 5 nm was formed on the $Ni_{0.2}Zn_{0.3}Fe_{2.5}O_4$ film by means of magnetron sputter deposition. As shown in Fig. 2(c), our TE sheet was highly flexible and easily bent without breaking the Pt/$Ni_{0.2}Zn_{0.3}Fe_{2.5}O_4$ film. The sheet was then cut into small pieces with a size of 8 × 4 $mm^2$ for TE measurements.

**Demonstration of the LSSE-based TE sheet for heat-flow sensing** To evaluate how well the LSSE-based TE sheet functioned as a heat-flow sensor, we investigated its TE property in the following fashion. A heat flux $q$ was driven across the 4×4-$mm^2$ central area of the TE-sheet sample by sandwiching the sheet between two Peltier modules. While driving the heat flow in such a manner, we simultaneously monitored the exact value of $q$ penetrating the TE-sheet sample with a commercially available thin-plate-shaped heat-flow sensor, which was set immediately above the sample. Because the commercial heat-flow sensor was placed in direct contact with the central area of the TE-sheet sample, we could assume that the heat flux value monitored by the sensor was the same as the $q$ actually penetrating across the sample. An external magnetic field $H$, which controls the direction of the magnetization **M** of the $Ni_{0.2}Zn_{0.3}Fe_{2.5}O_4$ films, was also applied to the entire system. The TE voltage $V$ between the two ends of the Pt film was measured with two contact probes. The resistance of the Pt film was determined to be $R_{Pt} = 238$ Ω.

Figure 2(d) represents $V$ as a function of $H$, measured when heat fluxes of $q$ = -13.7, -6.5, 0.0, 5.6, and 11.6 kW/m$^2$ were driven across the TE-sheet sample. The TE voltage was observed along the direction perpendicular to the direction of both $q$ and $H$, as derived from equation 1 (see the inset of Fig. 2(d)). The result shows that the sign of $V$ is flipped when $q$ or $H$ is reversed, which is a typical behaviour of LSSE-based devices. The heat-flux dependence of the TE voltage in Fig. 2(e) clearly demonstrates that $V$ is proportional to $q$. The heat-flow sensitivity derived from the fitted line is $V/q = 0.98$ nV/(W/m$^2$). The demonstration of this linear relationship between $V$ and $q$ suggests that our LSSE-based TE sheet functioned as a heat-flow sensor.

In an additional experiment, we have confirmed that the TE sheet exhibits no output signal when a temperature gradient was applied in the in-plane direction (see Supplementary Information). It suggests that the TSSE is negligibly small in our ferrite-plated film because of its 1D spin-current conducting property.

**Ferrite-thickness dependence of the LSSE-based TE sheet** We performed additional experiments to ascertain the origin of the observed TE signal. Given that a ferrite composed of $Ni_{0.2}Zn_{0.3}Fe_{2.5}O_4$ is typically a semiconducting ferrimagnet with a small but non-zero electrical conductivity, it can exhibit the ANE, which also produces a transverse voltage in the same experimental configuration as the LSSE. In our ferrite plated film, however, the in-plane electrical resistance of the $Ni_{0.2}Zn_{0.3}Fe_{2.5}O_4$ film was too high to be measured, which may be partly attributed to the vertically oriented grain boundaries of the columnar-structured film. Due to this transverse electric insulation, we could not observe any signals originating from the bulk ANE in the $Ni_{0.2}Zn_{0.3}Fe_{2.5}O_4$. However, there still remains a possibility that the TE signal includes ANE contribution caused by magnetic proximity effects at the Pt/$Ni_{0.2}Zn_{0.3}Fe_{2.5}O_4$ interface [35].

To shed light on such TE conversion mechanism, we investigated the TE properties of samples with varied ferrite-film thicknesses $t_F$. Figure 3 presents the ferrite-thickness dependence of the heat-flow sensitivity $(V/q)_{Norm}$ normalized to the sensitivity at $t_F = 500$ nm. The $(V/q)_{Norm}$ values monotonically increase for $t_F < 100$ nm, whereas the $t_F$ dependence of $V$ becomes saturated for $t_F > 100$ nm. The plots are well fitted to an exponential curve $(V/q)_{Norm} = 1 - \exp(-t_F/\lambda)$ with $\lambda = 71$ nm.

Similar to recent LSSE studies using yttrium iron garnet (YIG) films [36], this ferrite-thickness dependence is consistently explained according to the magnon-driven LSSE scenario [27,28,37,38], in which a certain ferrite thickness region (corresponding to the magnon-propagation length) below the Pt/ferrite interface effectively contributes to the voltage generation. On the other hand, the proximity-ANE scenario, which can occur at the Pt/$Ni_{0.2}Zn_{0.3}Fe_{2.5}O_4$ interface, is not able to explain this dependence. Thus, our finding suggests that the obtained signal originated mainly from the bulk magnon spin current driven by the LSSE. The result also suggests that our columnar-crystalline film possesses good spin-current-conduction properties suitable for LSSE-based sensors. Though it is beyond the scope of this work, such 1D spin-current conductors might have unconventional magnon-propagation properties which is different from that of 3D conductors, because magnon-scattering events can be altered in such confined structure. Control of magnon propagation in low-dimensional conductors will be an exciting research topic for future work from both academic and practical viewpoints.

**Bending-curvature dependence of the LSSE-based TE sheet**  Finally, we investigated heat-flow-sensing capability of the flexible LSSE-based TE sheet when the sheet was bent. Figure 4(a) and 4(b) depicts the $H$-dependence of the TE voltage $V$ for the same Pt/$Ni_{0.2}Zn_{0.3}Fe_{2.5}O_4$/polyimide sample when a heat flux $q$ was applied across the samples over a 20×20-mm$^2$ area under condition where the sample was flat or bent (with a radius of curvature of $r$ = 17 mm), respectively. The dependence of the heat-flow sensitivity, $V/q$, on the curvature $r^{-1}$ is presented in Fig. 4(c). The result clearly demonstrates that $V/q$ is nearly constant independent of $r^{-1}$, suggesting that the bending stresses applied to the Pt/$Ni_{0.2}Zn_{0.3}Fe_{2.5}O_4$ films do not significantly affect the TE conversion process consisting of the LSSE and the ISHE. This TE property, i.e., the TE conversion is independent of bending condition, is quite desirable for heat-flow sensing applications on various curved surfaces, because we are able to avoid additional calibration steps that depend on individual measuring objects with various surface curvature.

**Discussion**

We successfully demonstrated that an LSSE-based flexible TE sheet with 1D spin-current conducting film functions as a heat-flow sensor. The ferrite-thickness dependence of the TE voltage suggests that the TE signal was caused predominantly by the LSSE, which is consistent with other reports using Pt/$NiFe_2O_4$ [39]. The magnon-propagation length in our $Ni_{0.2}Zn_{0.3}Fe_{2.5}O_4$ film perpendicular to the film plane is approximately 71 nm. The TE sheet exhibit nearly identical heat-flow sensitivity regardless of bending curvature, suggesting that our columnar-crystalline film retains good 1D spin-current-conduction properties even when bent.

Though the heat-flow sensitivity $V/q$ of our TE sheet is not high in this stage, the outstanding features of it in contrast to currently available sensors are high flexibility in shape and remarkably low thermal resistance, which is a highly desirable feature for versatile heat-flow sensing. Although we formed ferrite films on plastic substrates in this work, it is also possible to directly plate various heat sources with ferrite films, thereby offering a thermal-sensing function while minimally obstructing the innate heat flux. Such features will offer a variety of opportunities for less destructive heat-flow measurements.

To use the TE sheets for a wide range of practical applications, the heat-flow sensitivity $V/q$ must be further improved. A straightforward method to enhance the $V/q$ is to enlarge the size of the TE sheet, as the output voltage scales linearly with the film length $l$. We can also increase the effective length $l$ inside a certain area by adopting meandering-patterned metallic film structures [20,40]. Another strategy is to replace Pt. We investigated TE-sheet samples with different metallic materials instead of Pt, and found that heat-flow sensitivity $V/q$ of W/$Ni_{0.2}Zn_{0.3}Fe_{2.5}O_4$ was 3.5-fold larger than that of Pt/$Ni_{0.2}Zn_{0.3}Fe_{2.5}O_4$ (see Supplementary Information). Moreover, we can enhance the sensitivity further by adopting recently reported (FI/NM)$_n$ multilayer structures [19], or FI/FM structure which can utilize both the SSE and ANE [20,21]. Although we applied an external magnetic field $H$ to the TE sheet for our experimental demonstration, this step is not necessary if the spontaneous magnetization **M** of the ferrite is sufficiently stable. The improvement of such magnetic stability is realized, for example, by doping cobalt into ferrite-plated films, which is known to enhance a coercive field of the ferrites [41]. The LSSE-based heat-flow-sensing technique, in which a heat flux induces an electrical signal indirectly via an LSSE-driven spin current, offers unconventional device-design opportunities, leading to novel heat-managing applications.

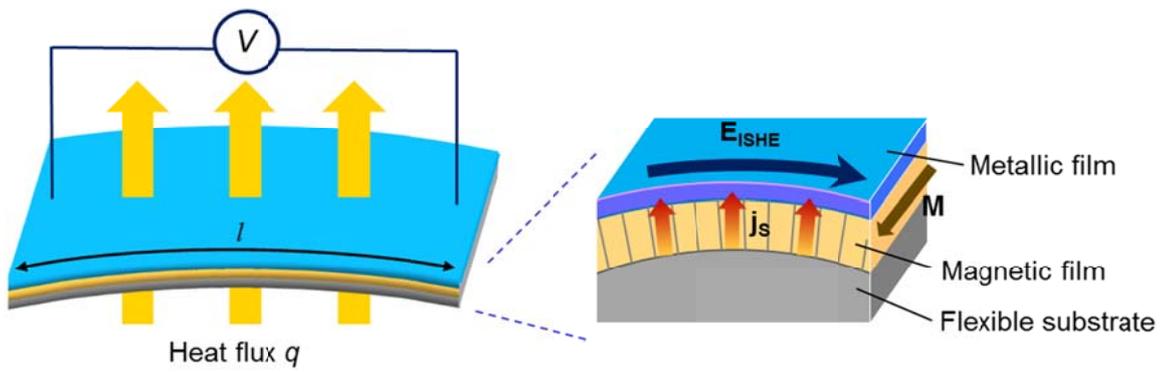

**Figure 1 | Concept of TE sheet for heat-flow sensing based on the LSSE**. The LSSE-based TE sheet consists of a metallic film and a magnetic (ferro- or ferrimagnetic) film formed on a flexible substrate. When a heat flux $q$ flows through the TE sheet, a spin current $\mathbf{j}_s$ is induced and injected from the ferrite film into the metallic film by the LSSE. Then the $\mathbf{j}_s$ is finally converted into an electric voltage $V$ as a result of the inverse spin Hall effect (ISHE) in the metallic film. The thin and simple bilayer structure of the TE sheet allows us to design novel heat-flow sensors with low thermal resistance and a flexible shape.

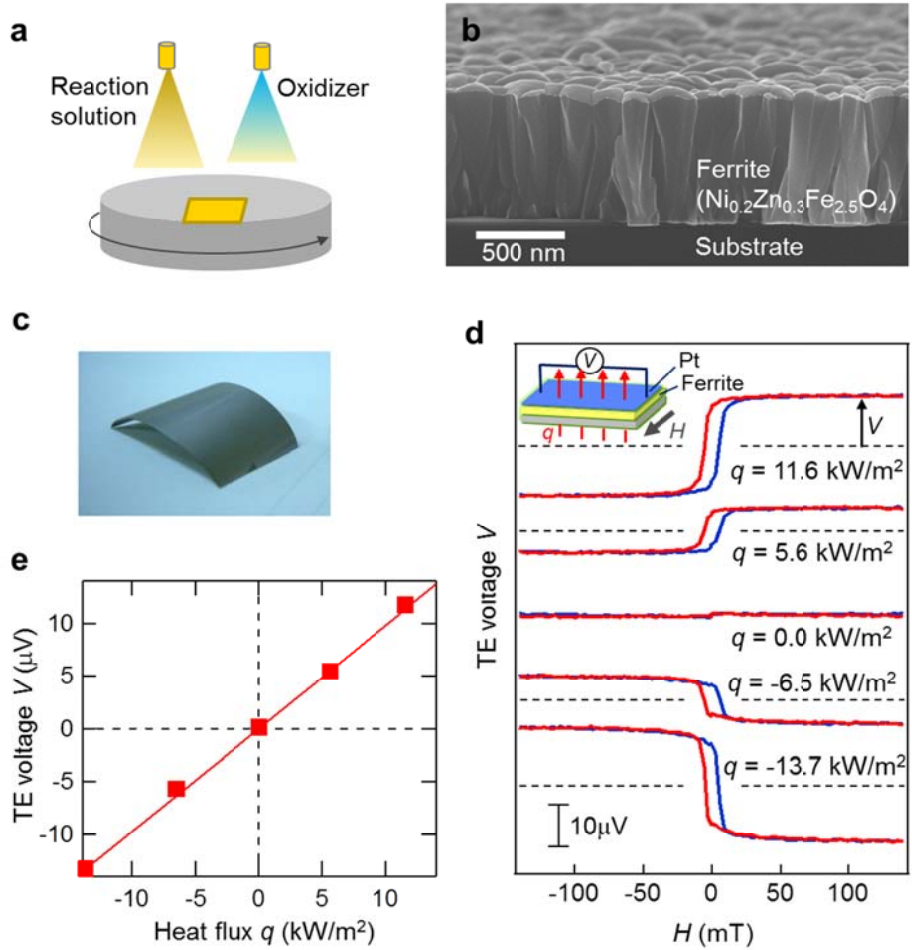

**Figure 2 | Demonstration of heat-flow-sensing TE sheet based on the LSSE.** (a) Schematic of the ferrite-plating method. An aqueous reaction solution ($FeCl_2+NiCl_2+ZnCl_2$) and an oxidizer ($NaNO_2+CH_3COONH_4$) are sprayed onto a substrate mounted on a rotating stage. (b) SEM image of a $Ni_{0.2}Zn_{0.3}Fe_{2.5}O_4$ film grown on a $SiO_2$/Si substrate using the ferrite-plating method. The film exhibits a columnar-crystal structure. The typical diameter of the columnar grains is approximately 100 nm. (c) Photograph of an LSSE-based flexible TE sheet, in which a Pt/$Ni_{0.2}Zn_{0.3}Fe_{2.5}O_4$ film was formed on a 25-μm-thick polyimide substrate. (d) TE voltage $V$ as a function of an external magnetic field $H$, measured when a heat flux $q$ was applied across the TE sheet. The sign of $V$ is reversed, when the sign of $H$ or $q$ changes. (e) TE voltage from the TE sheet as a function of $q$. From the fitting with the solid line, the heat-flow sensitivity of this TE sheet was $V/q = 0.98$ nV/(W/m$^2$).

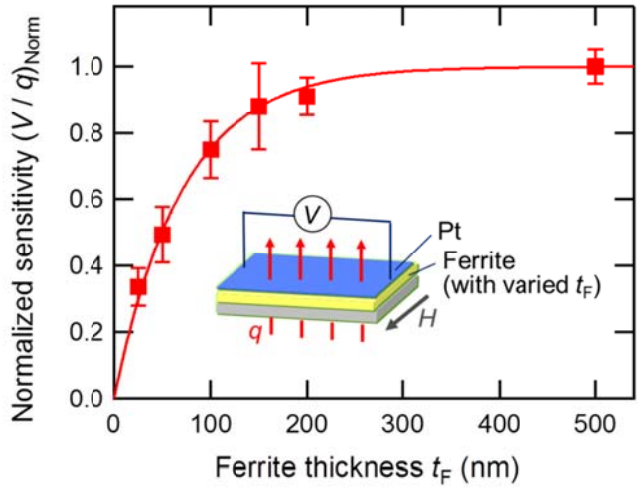

**Figure 3 | Ferrite-film-thickness dependence of LSSE-based TE sheets.** The $t_F$ dependence of the heat-flow sensitivity $V/q$, where the longitudinal axis is normalized by the sensitivity at $t_F = 500$ nm. The dependence is well fitted by an exponential curve $(V/q)_{Norm} = 1-\exp(-t_F/\lambda)$ with $\lambda = 71$ nm, consistent with the magnon-driven LSSE scenario.

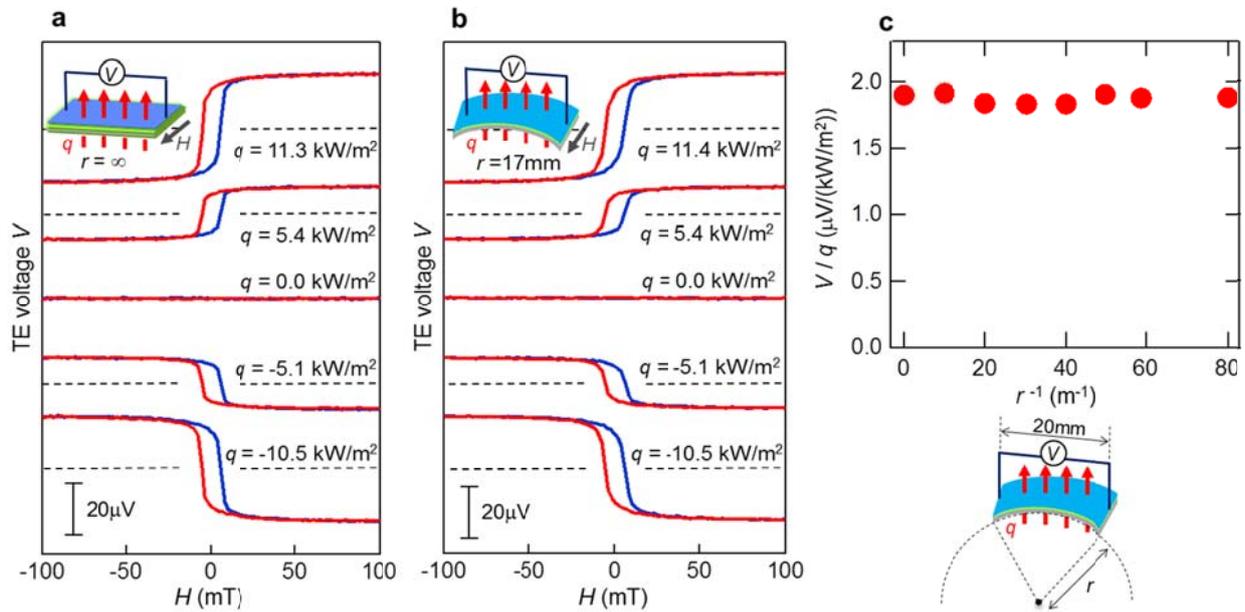

**Figure 4 | Heat-flow sensing with a bent TE sheet.** (a) TE voltage $V$ from a flat Pt/Ni$_{0.2}$Zn$_{0.3}$Fe$_{2.5}$O$_4$/polyimide sample as a function of an external magnetic field $H$ measured when the heat flux $q$ was applied across the sample over a 20×20-mm$^2$ area. (b) $H$ dependence of the TE voltage $V$ from a Pt/Ni$_{0.2}$Zn$_{0.3}$Fe$_{2.5}$O$_4$/polyimide sample that was bent with a radius of

curvature $r = 17$ mm, when the heat flux $q$ was applied across the sample over a 20×20-mm$^2$ area. (c) Heat-flow sensitivity $V/q$ as a function of curvature $r^{-1}$, indicating that $V/q$ is almost independent of $r^{-1}$.

**Methods**

**Sample preparation.** To prepare the $Ni_{0.2}Zn_{0.3}Fe_{2.5}O_4$ film for the TE sheet via a ferrite plating method, we first mounted a 25-μm-thick polyimide substrate on a rotating stage and then sprayed an aqueous reaction solution ($FeCl_2$+$NiCl_2$+$ZnCl_2$) and an oxidizer ($NaNO_2$+$CH_3COONH_4$) from two nozzles placed above the stage, as shown in Fig. 2a. This setup enabled us to grow the ferrite film by alternating adsorption and oxidation of the ingredient materials (including Fe, Ni, and Zn). During the process, the temperature of the stage was maintained at approximately 90 °C. The thickness of the $Ni_{0.2}Zn_{0.3}Fe_{2.5}O_4$ film was controlled via the time period of this formation process. The composition of the ferrite film was analysed by inductively coupled plasma spectroscopy (ICPS). A Pt film was deposited on the top of the $Ni_{0.2}Zn_{0.3}Fe_{2.5}O_4$ film with a magnetron sputtering system. Immediately before the sputtering process, the sample was exposed to argon plasma for 10 s to clean the surface of the $Ni_{0.2}Zn_{0.3}Fe_{2.5}O_4$.

**TE conversion measurements.** To evaluate the TE conversion of a heat flow to electric voltage, the sample was cut into small $8 \times 4$-mm$^2$ pieces using a cutter. To investigate the heat-flow-sensing properties of the LSSE-based TE sheet, we drove a heat flow across the sheet using two commercial $4 \times 4$-mm$^2$ Peltier modules. The two Peltier modules were attached to the top and bottom of the TE sheet, enabling us to heat one side and cool the other side of the TE sheet. The temperature difference, applied in such a manner, led to a heat flux penetrating through the TE sheet. Because the in-plane thermal conductance in our thin TE sheet was quite small, we can assume that the direction of the heat flux was nearly perpendicular to the TE sheet. While driving the heat flow, we simultaneously monitored the exact value of $q$ penetrating the TE sheet using a commercial thin-plate-shaped heat-flow sensor. The sensor was placed between the upper Peltier module and the TE sheet, in direct contact with the Pt film of the TE sheet. With this setup, we can assume that the same amount of heat flux $q$ flowed across both the TE sheet and the sensor. The generated TE voltage was measured with a digital multimeter.

**TE measurements of the bent samples.** To evaluate bent LSSE-based TE sheets as shown in Fig. 4, we used pairs of oxide-coated aluminium blocks with curved (concave and convex) surfaces. In the experiments, the TE sheet was sandwiched by the concave and convex blocks with a certain bending curvature. To investigate the bending-curvature dependence, we prepared several pairs of such blocks with different surface curvatures, in which the lateral size of the blocks was fixed to $20 \times 20$ mm$^2$. The heat-flow-sensing properties of the bent TE sheets were evaluated in the same manner as described above. The heat flux was driven across the TE sheet by two Peltier modules attached to the top and bottom of the block pair that sandwiched the sheet. Commercially available 20×20-mm$^2$ heat-flux sensors were also used to monitor the level of the heat flux penetrating across the TE sheet.

**Acknowledgements**

This work was partially supported by PRESTO "Phase Interfaces for Highly Efficient Energy Utilization" from JST, Japan, Grant-in-Aid for Scientific Research on Innovative Area, "Nano Spin Conversion Science" (No. 26103005), Grant-in-Aid for Challenging Exploratory Research (No. 26600067), and Grant-in-Aid for Scientific Research (A) (No. 15H02012) from MEXT, Japan.

**Author contributions**

A. K., M. I., K. U., E. S., S. K. and T. M. designed the experimental plan. A. K., K. K., H. S. and N. Y. mainly worked on sample preparation. A. K., M. I., K. I., Y. I. and A. M. performed the TE-conversion experiments. All the authors contributed to the analysis and discussion of the research.

**Additional information**

The authors declare no competing financial interests.

# Supplementary Information:
# Flexible heat-flow sensing sheets based on the longitudinal spin Seebeck effect using one-dimensional spin-current conducting films

Akihiro Kirihara[1,2*], Koichi Kondo[3], Masahiko Ishida[1,2], Kazuki Ihara[1,2], Yuma Iwasaki[1], Hiroko Someya[1,2], Asuka Matsuba[1], Ken-ichi Uchida[4,5], Eiji Saitoh[2,4,6,7], Naoharu Yamamoto[3], Shigeru Kohmoto[1] and Tomoo Murakami[1]

[1]*Smart Energy Research Laboratories, NEC Corporation, Tsukuba, 305-8501, Japan*

[2]*Spin Quantum Rectification Project, ERATO, Japan Science and Technology Agency, Sendai, 980-8577, Japan*

[3]*NEC TOKIN Corporation, Sendai, 982-8510, Japan*

[4]*Institute for Materials Research, Tohoku University, Sendai, 980-8577, Japan*

[5]*PRESTO, Japan Science and Technology Agency, Saitama, 332-0012, Japan*

[6]*WPI, Advanced Institute for Materials Research, Tohoku University, Sendai, 980-8577, Japan*

[7]*Advanced Science Research Center, Japan Atomic Energy Agency, Tokai, 319-1195, Japan*

**TE experiment in TSSE setup**   We have performed additional experiments to check whether transverse spin Seebeck effect (TSSE) [1-3] contributes to the TE signal in our devices. The schematics of the experimental setup is shown in Fig S1(a). For the TSSE experiment, we prepared a TE-sheet sample in a similar fashion as described in the main article. In the sample, a $20 \times 5$-mm$^2$ Pt strip with a thickness of 5 nm was formed on an edge of $20 \times 20$-mm$^2$ ferrite-plated film, as shown in Fig. S1(a). To investigate the TSSE, output voltage $V$ along the $y$-direction between two ends of the Pt strip was measured, when a temperature difference $\Delta T$ was applied in the $x$-direction. To magnetize the ferrite-plated film, an external magnetic field $H$ was also applied in the $x$-direction. If the TE-sheet sample exhibits the TSSE, output voltage is expected to occur in the $y$-direction. Figure S1(b) shows the measured $V$ as a function of $H$ when $\Delta T = 1.1$K was applied to the sample, where no output signal was clearly observed. The result suggests that our TE sheet with a ferrite-plated film having one-dimensional spin-current conducting properties does not exhibit any TE voltage originating from the TSSE, since a transverse spin current is effectively blocked by its columnar crystalline structure.

**LSSE-based TE sheets with different metallic films**   To gain further insights into the TE mechanism, we also prepared and evaluated TE sheets using different metal-film materials instead of Pt. TE measurements were performed in the same experimental configuration as mentioned above. Figure S2(a) represents the TE voltage from a sample composed of a 10-nm-thick Cu film and a 500-nm- thick $Ni_{0.2}Zn_{0.3}Fe_{2.5}O_4$ film on a polyimide substrate, showing that the output voltage from the

Cu film is negligibly small. This result is consistent with the negligible ISHE in Cu, which has a weak spin-orbit interaction. Figure S2(b) shows the experimental result of a TE sheet in which a W film with a thickness of 5 nm was deposited on the same $Ni_{0.2}Zn_{0.3}Fe_{2.5}O_4$/polyimide substrate. In this case, the clear TE voltage $V$ was observed and its sign was found to be opposite to that of the Pt/ $Ni_{0.2}Zn_{0.3}Fe_{2.5}O_4$ sample (compare Fig. S2(b) with Fig. 2(d) in the main article), which is consistent with the fact that the spin-Hall angle of W has a sign opposite to that of Pt [4,5]. Notably, the heat-flow sensitivity of the W/$Ni_{0.2}Zn_{0.3}Fe_{2.5}O_4$ sensor is $V/q = 3.55$ nV/(W/m$^2$), a value 3.5-fold greater than that of the Pt/$Ni_{0.2}Zn_{0.3}Fe_{2.5}O_4$, although the W-film resistance between the ends of the sample ($R_W = 2.04$ kΩ) was an order of magnitude greater than that of the Pt film. This large value suggests that W appears to be a promising material for heat-flow sensing applications.

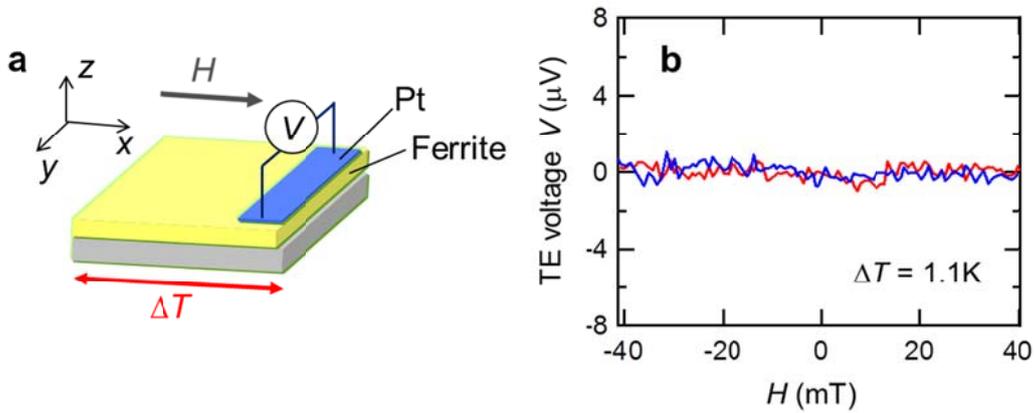

**Figure S1 | TE experiment in transverse-SSE setup.** (a) Experimental set up for checking the transverse SSE. Output voltage $V$ along the $y$-direction between two ends of the Pt strip was measured when temperature difference $\Delta T$ was applied in the $x$-direction. To magnetize the ferrite, an external magnetic field $H$ was also applied in the $x$-direction. (b) Measured voltage $V$ as a function of external magnetic field $H$ when $\Delta T = 1.1$K was applied to the sample.

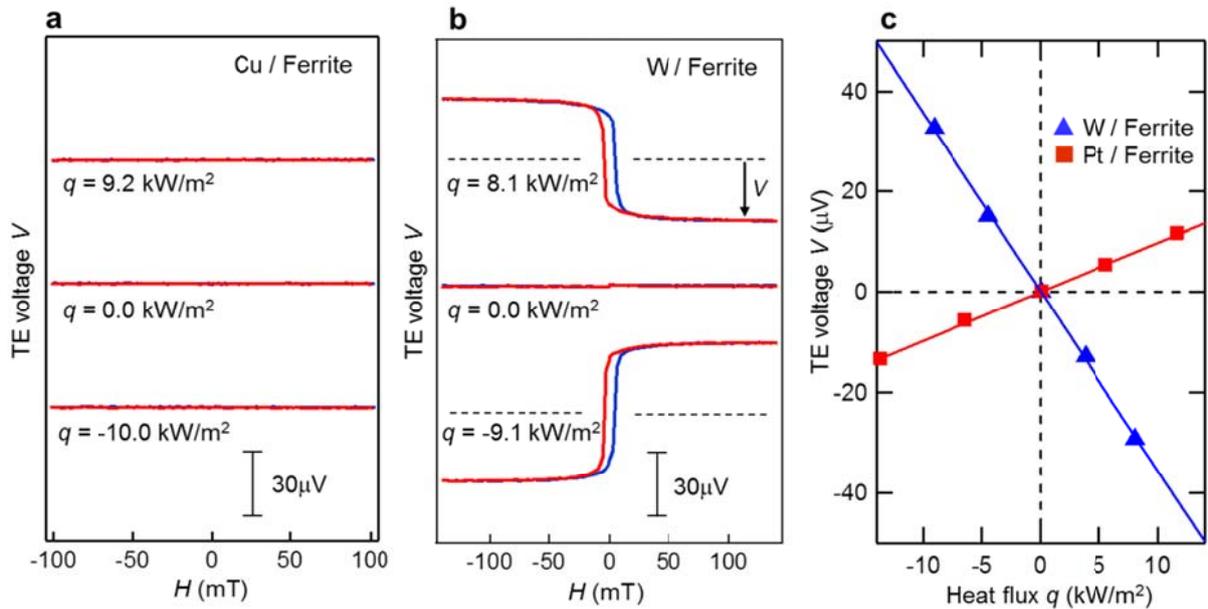

**Figure S2 | SSE-based TE sheets with different metallic films.** (a) TE voltage $V$ from a Cu/$Ni_{0.2}Zn_{0.3}Fe_{2.5}O_4$/polyimide sample as a function of an external magnetic field $H$, measured when the heat flux $q$ was applied across the sample. The voltage was negligibly small, due to the small ISHE in Cu. (b) TE voltage $V$ from a W/$Ni_{0.2}Zn_{0.3}Fe_{2.5}O_4$/polyimide sample obtained with the same measurement setup. The voltage signal has a sign opposite to that of the Pt/$Ni_{0.2}Zn_{0.3}Fe_{2.5}O_4$ sample

because the spin-Hall angle of W has the opposite sign to that of Pt. (c) TE voltage from the TE sheet W/Ni$_{0.2}$Zn$_{0.3}$Fe$_{2.5}$O$_4$ compared with that from Pt/Ni$_{0.2}$Zn$_{0.3}$Fe$_{2.5}$O$_4$ as a function of $q$. According to the fitting with the solid line, the heat-flow sensitivity of W/Ni$_{0.2}$Zn$_{0.3}$Fe$_{2.5}$O$_4$ was evaluated to be $V/q = 3.55$ nV/(W/m$^2$), which is more than 3 times larger than that of Pt/Ni$_{0.2}$Zn$_{0.3}$Fe$_{2.5}$O$_4$.